\documentstyle[amstex,amsfonts,epsfig,12pt]{article}

\begin{document}

\onecolumn

\begin{titlepage}
\begin{center}
\hfill{WATPHYS-TH00/06}
\\ \vspace{1.5cm}
{\LARGE \bf Exact Charged Two-Body Motion \\
 and the Static Balance Condition in Lineal Gravity}
\\ \vspace{2cm}
R.B. Mann \footnotemark\footnotetext{email: mann@@avatar.uwaterloo.ca} 
\\
\vspace{1cm}
Dept. of Physics,
University of Waterloo
Waterloo, ONT N2L 3G1, Canada\\
\vspace{2cm}
T. Ohta
\vspace{1cm}
Department of Physics, 
Miyagi University of Education,
Aoba-Aramaki, Sendai 980, Japan \\
\vspace{2cm}
\end{center}

\begin{abstract}
We find an exact solution to  the charged 
$2$-body problem in $(1+1)$ dimensional lineal gravity which
provides the first example of a relativistic system that generalizes 
the Majumdar-Papapetrou condition for static balance. 
\end{abstract}
\end{titlepage}\onecolumn

The static balance problem is a long-standing problem in gravitational
physics. Originally motivated by attempts to find exact solutions of the $%
(N\geq 2)$-body system in general relativity \cite{KSMH}, one seeks an
equilibrium solution in which gravitational attraction is balanced with
another repulsive force, typically electromagnetism. \ The first such
solution was found in Einstein-Maxwell theory by Majumdar \cite{Maj} and
Papapetrou \cite{Papa} \ (MJ) for $N=2$ and was later generalized to $N$
bodies on a line \cite{GHA}. The MJ static balance condition is 
\begin{equation}
e_{i}=\pm \sqrt{4\pi G}\;m_{i} \qquad (i=1, 2)  \label{M-P}
\end{equation}
and is considerably more stringent than the corresponding non-relativistic
condition 
\begin{equation}
Gm_{1}m_{2}-\frac{e_{1}e_{2}}{4\pi }=0\;.  \label{Newt-b}
\end{equation}

The reason why these conditions differ have long intrigued theorists. There
is no proof that (\ref{M-P}) is a necessary condition for static balance,
although it is sufficient. Although it has been conjectured \cite{Bonn} that
an exact solution under the condition (\ref{Newt-b}) should exist in general
relativity, in the (2nd) post-Newtonian approximation (\ref{Newt-b}) is
incompatible with the static balance condition \cite{OK}, and a test
particle analysis \cite{Bonn2} suggested that (\ref{Newt-b}) is neither
necessary nor sufficient. This suggests\ a wider range of possibilities for
realizing equilibrium in general relativity that do not exist
non-relativisitically, and several numerical studies \cite{Perry,Breton}
have been carried out to this end. \ Until now no one has yet found -- in
any relativistic theory of gravity -- equilibrium states in which $\sqrt{%
4\pi G}m_{i}>e_{i}$ for both bodies. \ 

We present in this paper a new equilibrium solution to the static balance
problem for which the condition (\ref{M-P}) does not hold. Our exact
solution is obtained for lineal gravity minimally coupled to
electromagnetism, and allows for the possibility that the masses of the
particles are both larger than their charges. This is the first example of
its type, and our full solution is the first non-perturbative relativistic
curved-spacetime treatment of this problem, providing new avenues for the
study of lineal self-gravitating systems. Indeed, one-dimensional
self-gravitating $N$-body systems have been quite fruitful in yielding
insight into many problems in gravity \cite{yawn}: they admit a considerably
simpler level of computational and analytic analysis that can be applied to
star systems (small $N\geq 2$) and galactic evolution (large $N$), whilst
avoiding a number of difficulties inherent in three dimensions, including
singularities, evaporation, and energy dissipation in the form of
gravitational radiation.

Our solution is derived in the context of the canonical theory for a charged 
$N$-body relativistic self-gravitating lineal system. \ We couple $N$
charged point masses to Jackiw-Teitelboim lineal gravity \cite{JT}, which in
the absence of matter equates the scalar curvature to a cosmological
constant 
\begin{equation}
R-\Lambda =0\;.  \label{RL}
\end{equation}
As a model theory of quantum gravity \cite{jtrfs} this model has been of
considerable interest; our modification to include charged particles yields
a generally covariant self-gravitating system with non-zero curvature
outside the point sources. We do not include collisional terms, so that the
bodies pass through each other.

We take the action to be 
\begin{eqnarray}
I &=&\int d^{2}x\left[ \frac{\sqrt{-g}}{2\kappa }g^{\mu \nu }\left\{ \Psi
R_{\mu \nu }+\frac{1}{2}\nabla _{\mu }\Psi \nabla _{\nu }\Psi +g_{\mu \nu
}\Lambda -\frac{\kappa }{2}F_{\mu}^{\;\;\alpha}F_{\nu \alpha }\right\}
\right.  \label{act1} \\
&&\left. -\sum_{a}\int d\tau _{a}\left\{ m_{a}\sqrt{\left( -\frac{dz_{a\mu }%
}{d\tau _{a}}\frac{dz_{a}^{\mu }}{d\tau _{a}}\right) }-e_{a}\frac{%
dz_{a}^{\mu }}{d\tau _{a}}A_{\mu }(x)\right\} \delta ^{\left( 2\right)
}(x-z_{a}(\tau _{a}))\right] \;,  \nonumber
\end{eqnarray}
where $\Psi $ is the dilaton field, which must be included since the
Einstein action is a topological invariant in 2 spacetime dimensions. Here $%
g_{\mu \nu }$ and $g$ are the metric and its determinant, $R$ is the Ricci
scalar, $\kappa =8\pi G/c^{4}$, and the electromagnetic field $F_{\mu \nu
}=\partial _{\mu }A_{\nu }-\partial _{\nu }A_{\mu }$, with $\tau _{a}$ the
proper time of $a$-th particle whose mass is $m_{a}$ and charge is $e_{a}$.
Variation of the action (\ref{act1}) with respect to the metric, dilaton
field, vector potential and particle coordinates yields the field equations 
\begin{equation}
R-\Lambda =\kappa T_{\;\;\mu }^{\mu },\;\frac{dz_{a}^{\alpha }}{d\tau _{a}}%
\nabla _{\alpha }\left\{ \frac{dz_{a}^{\nu }}{d\tau _{a}}\right\} =\frac{%
e_{a}}{m_{a}}\frac{dz_{a}^{\alpha }}{d\tau _{a}}F_{\,\,\alpha }^{\nu
}(z_{a})\;,  \label{RTgeo}
\end{equation}
\begin{equation}
\partial _{\nu }\left( \sqrt{-g}F_{\;\;}^{\mu \nu }\right)
=\sum_{a}e_{a}\int d\tau _{a}\frac{dz_{a}^{\mu }}{d\tau _{a}}\delta
^{2}(x-z_{a}(\tau _{a}))\;,  \label{RTem}
\end{equation}
\begin{eqnarray}
\frac{1}{2}\nabla _{\mu }\Psi \nabla _{\nu }\Psi -g_{\mu \nu }\left( \frac{1%
}{4}\nabla ^{\lambda }\Psi \nabla _{\lambda }\Psi -\nabla ^{2}\Psi \right)
-\nabla _{\mu }\nabla _{\nu }\Psi &&  \nonumber \\
=\kappa T_{\mu \nu }+\frac{\Lambda }{2}g_{\mu \nu }\;, &&  \label{E4}
\end{eqnarray}
where the stress-energy is due to the electromagnetic field and the point
masses 
\begin{eqnarray*}
T_{\mu \nu } &=&\sum_{a=1}^{N}m_{a}\int \frac{d\tau _{a}}{\sqrt{-g}}g_{\mu
\sigma }g_{\nu \rho }\frac{dz_{a}^{\sigma }}{d\tau _{a}}\frac{dz_{a}^{\rho }%
}{d\tau _{a}}\delta ^{\left( 2\right) }(x-z_{a}(\tau _{a})) \\
&&+\left\{ F_{\mu \alpha }F_{\nu }^{\;\alpha }-\frac{1}{4}g_{\mu \nu
}F_{\alpha \beta }F^{\alpha \beta }\right\}
\end{eqnarray*}
and is conserved. The set (\ref{RTgeo},\ref{RTem}) is a closed system of $%
N+2 $ equations whose solution yields the single metric and electromagnetic
degrees of freedom and the $N$ degrees of freedom of the point masses; it
reduces to (\ref{RL}) if all the masses $m_{a}$ and the charges $e_{a}$
vanish. Note that the evolution of the charged point-masses governs that of
the dilaton field via (\ref{E4}). The left-hand side of (\ref{E4}) is
divergenceless (consistent with the conservation of $T_{\mu \nu }$),
yielding only one independent equation to determine the single degree of
freedom of the dilaton.

Working in the canonical formalism where $\gamma
=g_{11},N_{0}=(-g^{00})^{-1/2},N_{1}=g_{10}$, and $A_{\mu }=(-\varphi ,A)$,
we have been able to extend the exact solution we previously obtained for
neutral bodies \cite{prl} to the charged case. After elimination of
coordinate and gauge degrees of freedom using standard methods \cite{adm},
the only independent degrees of freedom are the momenta $p_{i}$\ and spatial
coordinates $z_{i}$\ of the particles. \ In the two particle case these
reduce to $r\equiv z_{1}-z_{2}$ and $p_{1}=-p_{2}=p$, and the Hamiltonian is
determined from the equation

\begin{equation}
\tanh (\frac{\kappa {\cal J}}{8}|r|)=\frac{{\cal J}(B_{1}+B_{2})}{{\cal J}%
^{2}+B_{1}B_{2}}\;,  \label{defineH}
\end{equation}
where ${\cal J}^{2}=\left( \sqrt{H^{2}+\frac{8\Lambda _{e}}{\kappa ^{2}}}%
-2\epsilon \tilde{p}\right) ^{2}-\frac{8e_{1}e_{2}}{\kappa }-\frac{8\Lambda
_{e}}{\kappa ^{2}},B_{1,2}=H-2\sqrt{p^{2}+m_{1,2}^{2}}$, $\tilde{p}_{i}=p_{i}%
\mbox{sgn}(z_{1}-z_{2})$,\ and $r\equiv z_{1}-z_{2}$. The parameter $%
\epsilon =\pm 1$ is a constant of integration associated with the metric
degree of freedom and changes sign under time reversal. We have written $%
\Lambda _{e}\equiv \Lambda -\frac{\kappa }{4}(\sum_{a}e_{a})^{2}$, which is
an effective cosmological constant in the spacetime. When $\Lambda _{e}=0,$%
the Hamiltonian $H$ reduces to that of two charged point particles in the
non-relativistic limit \cite{ohtarobb}, containing that in ref. \cite{yawn}
when $e_{a}=0$. For a given $\Lambda _{e}\geq -(\kappa H)^{2}/8$, the
equation (\ref{defineH}) describes the surface in $(r,p,H)$ space of all
allowed phase-space trajectories. A given trajectory in the $(r,p)$ plane is
uniquely determined by setting $H=H_{0}$ in (\ref{defineH}), since $H$ is a
constant of the motion (a fact easily verified by differentiation of (\ref
{defineH}) with respect to $t$). \ 

The explicit solution for the field components -- although formally the same
as that obtained in ref. \cite{prl} -- is rather complicated, and will be
omitted here. \ However the equations of motion for $\dot{p}_{a}$ and $\dot{%
z_{a}}$ from (\ref{RTgeo}) have an additional Lorentz-force term which
yields qualitatively new features. In the 2-body unequal mass case they can
be transformed in terms of a new time coordinate to integral form. \ These
integrals cannot be computed in terms of elementary functions except in the
equal mass case where the exact solution is 
\begin{equation}
p(\tau )=\frac{\epsilon m}{2}\left( f(\tau )-\frac{1}{f(\tau )}\right) \;,
\label{p-exact1}
\end{equation}
with 
\[
f(\tau )= \left\{
\begin{array}{cc}
\frac{ \frac{H}{m} \left(1+\sqrt{\gamma_{H}}\right) 
\left\{ 1-\eta e^{\frac{\epsilon \kappa m}{4}
\sqrt{\gamma_{m}}(\tau -\tau_{0})} \right\} }
{\gamma_{e}+\sqrt{\gamma_{m}} + \left( \sqrt{\gamma_{m}}-\gamma _{e} \right)
\eta e^{\frac{\epsilon \kappa m}{4}\sqrt{\gamma_{m}}(\tau -\tau _{0}) } }
& \quad \gamma_{m}>0,  \\
\frac{1+\sqrt{\gamma_H}}{\frac{m}{H}\gamma_{e}
+ \sigma\left(m-\sigma\frac{\epsilon \kappa H}{8}(\tau -\tau_0) \right)^{-1}}   
& \quad \gamma_{m}=0, \\ 
\frac{H}{m}(1+\sqrt{\gamma_{H}})
\left[\gamma_{e}+\sqrt{-\gamma_{m}}
\frac{\sigma + \frac{m^{2}}{H}\sqrt{-\gamma_{m}}\tan\left[ 
\frac{\epsilon\kappa m}{8}\sqrt{-\gamma_{m}}(\tau -\tau _{0})\right]}
{\frac{m^{2}}{H}\sqrt{-\gamma_{m}}- \sigma \tan\left[ 
\frac{\epsilon \kappa m}{8} \sqrt{-\gamma_{m}}(\tau -\tau _{0})\right]}\right]^{-1} 
& \quad \gamma_{m}<0 ,
\end{array}
\right.
\]
where $d\tau =d\tau _{1}=d\tau _{2}=\frac{m}{\sqrt{p^{2}+m^{2}}}\frac{{\cal J%
}^{2}}{C}dt$ is the proper time of each particle and 
\begin{eqnarray*}
\gamma _{H} &=&1+\frac{8\Lambda _{e}}{\kappa ^{2}H^{2}}\;,\text{ \ \ \ }%
\gamma _{e}=1+\frac{2e_{1}e_{2}}{\kappa m^{2}}\;, \\
\gamma _{m} &=&\gamma _{e}^{2}+\frac{8\Lambda _{e}}{\kappa ^{2}m^{2}}\;,%
\text{ \ }\;\eta =\frac{\sigma -\frac{m^{2}}{H}\sqrt{\gamma _{m}}}{\sigma +%
\frac{m^{2}}{H}\sqrt{\gamma _{m}}}\;, \\
\text{\ }\sigma &=&(1+\sqrt{\gamma _{H}})(\sqrt{p_{0}^{2}+m^{2}}-\epsilon
p_{0})-\frac{m^{2}}{H}\gamma _{e}\;, \\
C &=&{\cal J}^{2}-(H-\frac{2\epsilon \tilde{p}}{\sqrt{\gamma _{H}}})\left\{
B+\frac{\kappa }{16}({\cal J}^{2}-B^{2})r\right\} \;,
\end{eqnarray*}
with $p_{0}$ the initial momentum at $\tau =\tau _{0}$. It is then
straightforward to obtain an exact expression for $r$ as a function of $\tau 
$ by inserting the expression for $p(\tau )$ into (\ref{defineH}) and
solving for $r$ as a function of $\tau $.

\bigskip 
\begin{figure}[tbp]
\begin{center}
\epsfig{file=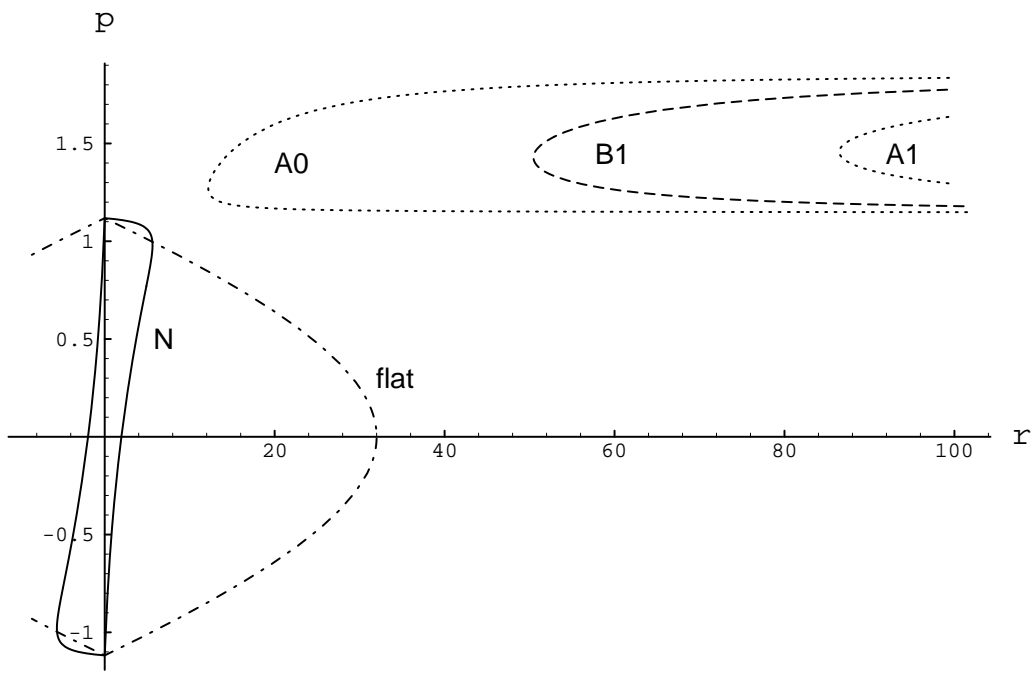,width=0.8\linewidth}
\end{center}
\caption{Phase space trajectories of unbounded motions for $H_{0}<2m$ 
($m=1$, $H_{0}=1$ and $e_{1}=e_{2}=\pm 1$).  The $\kappa =0$ limit is marked 
{\it ``flat''}.}
\label{fig1}
\end{figure}

\bigskip 
\begin{figure}[tbp]
\begin{center}
\epsfig{file=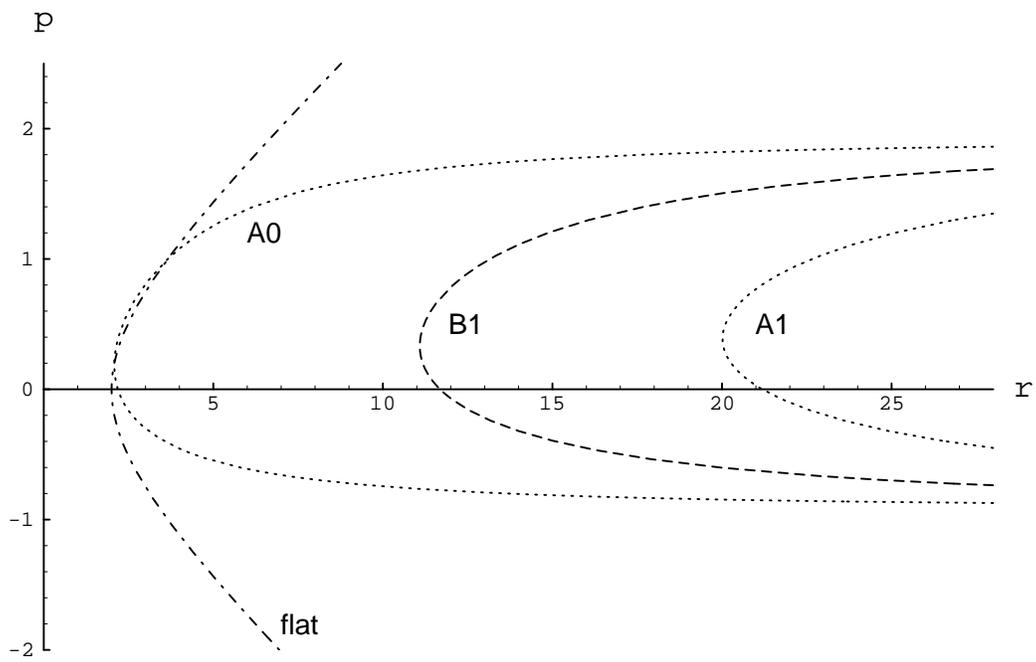,width=0.8\linewidth}
\end{center}
\caption{Phase space trajectories of bounded and unbounded motions for 
$H_{0}=3,m=1$ and $e_{1}=e_{2}=\pm 0.25$ in the repulsive case. The 
$\kappa =0$ limit is marked {\it ``flat''}.}
\label{fig2}
\end{figure}

A number of different types of motion are possible, depending upon a
combination of four factors: gravitational attraction, the electric force
between charges, the effect of the cosmological constant and relativistic
effects. The solution is characterized by the signs of $\gamma _{m}$ and $%
\kappa ^{2}\left( H-mf(\tau )+\frac{m}{f(\tau )}\right) ^{2}-8\kappa
e_{1}e_{2}-8\Lambda _{e}$. \ 

One of the qualitatively new features which arises is the existence of
unbounded motion for $H<2m$, provided the electromagnetic repulsion between
the charges is sufficiently strong. In Fig. 1 we plot (for $\Lambda _{e}=0$)
several phase space trajectories for $H=m$ and $|e_{1}e_{2}|=m(\kappa
=1,\epsilon =1)$, comparing with the analogous flat-space trajectory (a
dot-dashed curve). The additional effect of gravitational attraction causes
all trajectories to curve more toward the $r$-axis and to shift in the
direction of the positive $p$-axis due to the $p$-and $\epsilon $-
dependence of the gravitational potential.

More generally there will only be bounded motion whenever both $%
e_{1}e_{2}\leq 0$ and$\ \Lambda _{e}\leq 0$ in which the electromagnetic and
cosmological interactions are attractive, and only unbounded motion whenever
these inequalities are reversed. \ Otherwise both bounded and unbounded
motions will be present.\ Fig.1 is typical for all unbounded motions, and a
countably infinite series of unbounded trajectories exists for a fixed value
of the total energy $H$. Fig. 2 illustrates a phase space diagram in the $%
\Lambda _{e}=0$ repulsive case.

Our solution provides the first example of a new equilibrium solution to the
static balance problem. From the relation (\ref{defineH}) we compute the
balance condition $\partial H/\partial r=0$. This yields the relations 
\begin{equation}
{\cal J}(B_{1}+B_{2})={\cal J}^{2}+B_{1}B_{2}\;=0
\end{equation}
which in turn imply that $B_{1}=-B_{2}$ and ${\cal J}^{2}=B_{1}^{2}$. These
simplify to the condition

\begin{equation}
\frac{\kappa }{2}(\sqrt{p^{2}+m_{1}^{2}}-\epsilon \tilde{p})(\sqrt{%
p^{2}+m_{2}^{2}}-\epsilon \tilde{p})-e_{1}e_{2}=0  \label{bal-p}
\end{equation}
which we refer to as the force-balance condition. Only for $e_{1}e_{2}>0$ is
the value of the momentum fixed 
\begin{equation}
p=p_{c}{\bf =\pm }\frac{|\left( \frac{\kappa }{2}\right)
^{2}m_{1}^{2}m_{2}^{2}-e_{1}^{2}e_{2}^{2}|}{\sqrt{2\kappa e_{1}e_{2}}\sqrt{(%
\frac{\kappa }{2}m_{1}^{2}+e_{1}e_{2})(\frac{\kappa }{2}m_{2}^{2}+e_{1}e_{2})%
}}\;,  \label{pc}
\end{equation}
in which case the two particles move with constant velocity. Eq. (\ref{bal-p}%
) can also be inferred by perturbatively solving for $H$ in terms of $r$\
from (\ref{defineH}). When the particles are initially at rest $(p_{c}=0)$,
the condition (\ref{bal-p}) becomes 
\begin{equation}
\frac{\kappa }{2}m_{1}m_{2}-e_{1}e_{2}=0\;.  \label{relstat}
\end{equation}
which is the static balance condition, identical to the non-relativistic
condition (\ref{Newt-b}) ( also valid in (1+1) dimensions).

The condition (\ref{Newt-b}) applies to both static and uniform motion,
whereas the relativistic condition (\ref{relstat}) represents only a static
balance condition. The condition of force-balance (\ref{bal-p}) in general
depends on the momentum and can be satisfied for some fixed momentum $p_{c}$
whilst maintaining for both particles $\sqrt{4\pi G}m_{i}>e_{i}$ (or
alternatively $\sqrt{4\pi G}m_{i}<e_{i}$). This is a qualitatively new
feature and suggests that analogous equilibrium states might also exist in
(3+1) dimensional general relativity. This is not without precedent: in
(3+1) dimensions, for example, a Newtonian theory of gravitating charged
particles, if corrected to include the Darwin potential, has an analogous
equilibrium solution of constant momentum. It is an interesting open
question as to whether or not this feature will survive in a full
relativistic theory.

\bigskip

This work was supported in part by the Natural Sciences and Engineering
Research Council of Canada.

\end{document}